\documentclass{article}
\usepackage{graphicx}
\usepackage{epstopdf}

\usepackage{amsmath}
\usepackage{color}

\newcommand{\be}{\begin{equation}}
\newcommand{\ee}{\end{equation}}
\newcommand{\ben}{\begin{eqnarray}}
\newcommand{\een}{\end{eqnarray}}

\begin{document}

\markboth{Leandro Cesar Mehret and Gilberto Medeiros Kremer}
{Temperature oscillations of a gas  in Reissner-Nordstr\"om spacetime}

\title{Temperature oscillations of a gas moving close to circular geodesic in Reissner-Nordstr\"om spacetime  }

\author{Leandro C. Mehret\footnote{lcmehret@gmail.com} and  Gilberto M. Kremer\footnote{kremer@fisica.ufpr.br}\\ Departamento de F\'isica, Universidade Federal do Paran\'a \\ Caixa Postal 19044, 81531-980 Curitiba, Brazil}

\date{}
\maketitle

\begin{abstract}
The objective of this work is to analyze the temperature oscillations that occur in a gas in a circular motion under the action of a Reissner-Nordstr\"om gravitational field, verifying the effect of the charge term of the metric on the oscillations. The expression for temperature oscillations follows from Tolman's law written in Fermi normal coordinates for a comoving observer. The motion of the gas is close to geodesic so the equation of geodesic deviation was used to obtain the expression for temperature oscillations. Then these oscillations are calculated for some compact stars, quark stars, black holes and white dwarfs, using values of electric charge and mass from models found in the literature. Comparing the various models analyzed, it is possible to verify that the role of the charge is the opposite of the mass. While the increase of the mass produces a reduction in the frequencies, amplitude and in the ratio between the frequencies, the increase of the electric charge produces the inverse effect. In addition, it is shown that if the electric charge is  proportional to the mass, the ratio between the frequencies does not depend on the mass, but only on the proportionality factor between charge and mass. The ratios between the frequencies for all the models analyzed (except for supermassive black holes in the extreme limit situations) are close to the $3/2$ ratio for twin peak quasi-periodic oscillation frequencies, observed in many galactic black holes and neutron star sources in low-mass X-ray binaries.
\end{abstract}

%\tableofcontents

\section{Introduction}
\label{sec01}	

The Reissner-Nordstr\"om metric is a solution of Einstein's field equations that corresponds to the gravitational field produced by an electrically charged massive object (see e.g. the books \cite{wald,misner,chandrasekhar}). Some theoretical models use the idea of electrically charged astronomical objects to explain certain observable phenomena \cite{ray03,binnun11, zakharov14, liu14}, some of them involving compact objects such as neutron stars, quark stars, white dwarfs and black holes.

This work is carried out with the main purpose of determining the influence of the electric charge term in the Reissner-Nordstr\"om metric on the temperature oscillations, following the same methodology as adopted in ref. \cite{zimdahl15}. To reach this objective, we analyzed temperature oscillations for some theoretical models of charged compact objects found in the literature \cite{ray03, binnun11, zakharov14, liu14, negreiros09}. The problem consists in a gas of particles in circular motion around a charged massive object that produces a strong gravitational field. The motion is approximately geodesic, so the geodesic deviation equation was used to describe it. Throughout the article the convention $ (-, +, +, +) $ for the components of the metric tensor is used.

We consider a rarefied gas inside a spacecraft in a circular orbit around a charged massive object where the center of mass of the gas is moving on a circular orbit with geodesic deviation of the Reissner-Nordstr\"om metric. This is a relativistic gas so it can be described using Boltzmann equation and, considering the equilibrium, Maxwell-J\"uttner distribution function \cite{cercignani, zimdahl15}. From this, Tolman's law $\sqrt{-g_{00}}T=$ constant can be obtained, establishing a relation between the temperature and the geometry of the spacetime in equilibrium \cite{tolman30,klein}. This description using a Boltzmann gas at equilibrium is an idealized toy model, with several limitations but perhaps can illustrate some features of a real situation.

In the present study, we obtained the proper frequencies of the motion by analyzing the Lagrangian of a test particle of the gas in Reissner-Nordstr\"om metric. These frequencies were obtained in order to compare them with the frequencies calculated considering the geodesic deviation approximation for the motion of the gas. The orbital motion of a test particle around compact objects (in particular, the black hole in the galactic center) is subject for several recent publications \cite{hees17, broderick13, zhang15}. 
 
Because of the nature of the geodesic motion, it will be convenient to use Fermi normal coordinates. These coordinates make the Christoffel symbols vanish along the geodesic, leaving the metric locally rectangular \cite{manasse63}. The equilibrium condition can be applied to an approximately geodesic motion by the use of the geodesic deviation equation, since the perturbation terms are linear in distance. The geodesic deviation equation leads to the same frequencies obtained in the orbital motion analysis. The methodology applied in the present study was already used to describe other physical situations, as for example to calculate gravitational perturbation of the hydrogen spectrum \cite{parker82}, the Shirokov effect for sattelite orbits \cite{shirokov73}, and gravitationally induced supercurrents related to this effect \cite{zimdahl85}.

The phenomena of QPOs (quasi-periodic oscillations) are present in many galactic black holes and neutron star sources in low-mass X-ray binaries \cite{torok05}. The frequencies of some QPOs are in the $kHz$ range, corresponding to orbital frequencies next from the central black hole or neutron star. These are known as HF (high frequency) QPOs. These oscillations often show up in pairs, named twin peak HF QPOs \cite{germana13}. These two peaks correspond to different frequencies: the upper peak is assumed to be the modulation of the azimuthal frequency, whereas the lower one is a precession frequency. Typically, the ratio between these two frequencies is equal to $3/2$.

This $3/2$ ratio, along with other general properties of the double peak QPOs, are supposed to reflect a non-linear resonance between epicyclic oscillations in accretion fluid flows in super-strong gravity \cite{kluzniak01, kluzniak04}. These fluid accretion flows admit two quasi-incompressible modes of oscillations, vertical and radial, with corresponding eigenfrequencies equal to vertical and radial epicyclic frequencies for free particles \cite{torok05}. In the present study we found two frequencies for orbital motion in the $kHz$ range and the ratio of them is compared with the $3/2$ ratio found in QPOs. 

This article is organized as follows: section \ref{sec01a} presents the thermodynamical fundamentals needed for the analysis of the temperature oscillations.  In section \ref{sec02}, the methodology applied in this study is presented, obtaining the proper frequencies of the particle motion in Reissner-Nordstr\"om metric. The gas in a circular motion with geodesic deviation in Reissner-Nordstr\"om space-time is analyzed in section \ref{sec02a} where the Fermi normal coordinates for a comoving observer is introduced and the equations for the geodesic deviation are determined. The results are presented and discussed in \ref{sec03}, with the calculation of the temperature oscillations from Tolman's law for some models of electrically charged compact objects obtained in the literature. Section \ref{sec04} contains the conclusions regarding this research.

\section{Kinetic theory and Tolman's law}
\label{sec01a}

Consider a relativistic gas of particles with rest mass $m$ in a spacetime with metric tensor $g_{\mu \nu}$. The mass  shell condition $g_{\mu \nu} p^\mu p^\nu = -m^2 c^2$ allows us to describe the state of this gas by the one-particle distribution function $f(\vec{x},\vec{p},t)$ in the phase space spanned by the spacetime coordinates $(x^\mu) = (ct, \vec{x})$ and the momenta $(p^\mu) = (p^0, \vec{p})$ \cite{zimdahl15}. The evolution of the distribution function in the phase space is given by the Boltzmann equation \cite{cercignani} 
\begin{equation}
p^\mu \frac{\partial f}{\partial x^\mu} - \Gamma_{\mu \nu}^\sigma p^\mu p^\nu \frac{\partial f}{\partial p^\sigma} = Q(f,f),
\label{eq86}
\end{equation}
where $\Gamma_{\mu \nu}^\sigma$ are the Christoffel symbols and $Q(f,f)$ the collision operator. In equilibrium, the distribution function becomes the Maxwell-J\"uttner distribution function 
\begin{equation}
f^{(0)} = \frac{n}{4 \pi k T m^2 c K_2(\zeta)} \exp\left(\frac{U^\tau p_\tau}{kT}\right).
\label{eq87}
\end{equation}
Here $n$ is the particle number density, $T$ is the temperature, $U^\tau$ is the four-velocity, $k$ is the Boltzmann constant. $K_n(\zeta)$ denotes modified Bessel functions of the second kind with $\zeta = {mc^2}/{kT}$, given explicitly by 
\begin{equation}
K_n(\zeta) = \left(\frac{\zeta}{2}\right)^n \frac{\Gamma(1/2)}{\Gamma(n+1/2)} \int_{1}^{\infty} \exp(-\zeta y) (y^2-1)^{n-\frac12} dy
\label{eq88}
\end{equation}

Using the equilibrium distribution function (\ref{eq87}) we can calculate the energy-momentum tensor
\begin{equation}
T^{\mu \nu} = c \int p^\mu p^\nu f^{(0)}\sqrt{-g}\frac{d^3 p}{p^0} = \left(\frac{en + p}{c^2}\right)  U^\mu U^\nu + pg^{\mu \nu}
\label{eq89}
\end{equation}
In the equation (\ref{eq89}), $g$ denotes the determinant of the metric tensor, and the energy per particle $e$ and pressure $p$ are given by
\begin{equation}
e = mc^2 \left[ \frac{K_3(\zeta)}{K_2(\zeta)} - \frac{1}{\zeta}\right], \qquad p=nkT.
\label{eq89a}
\end{equation}

In equilibrium the entropy-flow vector reduces to 
\begin{equation}
S^{\mu} = -kc \int p^\mu f^{(0)} \ln f^{(0)}\sqrt{-g}\frac{d^3 p}{p^0} = nsU^\mu
\label{eq94}
\end{equation}
where $s$ denotes the entropy per particle 
\begin{equation}
s = k \left[\ln \left(\frac{4 \pi kTm^2 c K_2 (\zeta)}{n}\right)+ \zeta\frac{K_3(\zeta)}{K_2(\zeta)} - 1\right].
\label{eq94a}
\end{equation}

The  chemical potential $\mu$ is associated with the Gibbs function per particle $\mu = e-Ts+ p/n$ and it is given by
\begin{equation}
\mu  = kT\left[1+ \ln\frac{n}{4 \pi kTm^2 c K_2(\zeta)}\right].
\label{eq95}
\end{equation}

Now using (\ref{eq95}) the equilibrium distribution function (\ref{eq87}) can be written as 
\begin{equation}
f^{(0)} = \exp \left(\frac{\mu}{kT} -1 + \frac{U^\tau p_\tau}{kT}\right).
\label{eq96}
\end{equation}

The equilibrium distribution function (\ref{eq87}) is obtained from the condition that at equilibrium  the collision operator $Q(f,f)$ vanishes  (for more details see e.g. \cite{cercignani}). If we insert (\ref{eq96}) into the left-hand side of Boltzmann equation (\ref{eq86}) results 
\begin{equation}
p^\nu \partial_\nu\left[\frac{\mu}{kT}\right] - \frac{1}{2}p^\mu p^\nu \left(\left[\frac{U_\nu}{kT}\right]_{;\mu} + \left[\frac{U_\mu}{kT}\right]_{;\nu} \right) =0.
\label{eq105}
\end{equation}
The expression (\ref{eq105}) is valid for all $p^\mu$ so
\begin{equation}
\partial_\nu\left[\frac{\mu}{kT}\right]=0,\qquad \left[\frac{U_\nu}{kT}\right]_{;\mu} + \left[\frac{U_\mu}{kT}\right]_{;\nu}=0.
\label{eq106}
\end{equation}
The right-hand expression of the above equation is known as Killing equation, and ${U_\nu}/{kT}$ is a (timelike) Killing vector. This expression can be rewritten as
\begin{gather}
[U_\mu T^{-1}]_{;\nu}+ [U_\nu T^{-1}]_{;\mu} = U_{\mu;\nu}+ U_{\nu;\mu} - T^{-1}(U_\mu T_{,\nu}+ U_\nu T_{,\mu})=0
\label{eq107}
\end{gather}

By choosing suitable projections proportional and perpendicular to the four-vector $U^{\alpha}$ we get from (\ref{eq107})
\begin{equation}
\dot{T} = 0, \qquad \dot{U}_\mu + \frac{c^2}{T} \nabla_\mu T=0.
\label{eq112}
\end{equation}
In the above relations we used the definitions
$
\dot{T} \equiv U^\mu \partial_\mu T,$  $ \dot{U}_\mu \equiv U^\nu U_{\mu;\nu},$ $\nabla_\mu T \equiv h^\nu_\mu T_{,\nu}$
with $h_{\mu\nu} = g_{\mu\nu} + c^{-2}U_\mu U_\nu$ denoting the projector. 

The interpretation of equations (\ref{eq112}) is that at equilibrium a gas must have a stationary temperature and its acceleration must be counterbalanced by a spatial temperature gradient \cite{zimdahl15}. Note that there are situations where equation (\ref{eq112}) is compatible with geodesic motion, as for example the case for equilibrium tori \cite{kovar11}, where even if temperature is not uniform within the gas, there may be one world line on which it reaches a maximum value, and the world line would be geodesic. However, for a rarefied gas in circular motion around a charged massive object with strong gravitational field, the right-hand expression of condition (\ref{eq112}) is not compatible with a geodesic fluid motion, which would require $\dot{U}^\mu = 0$. In section \ref{sec02b} we shall return to this point. 

We can obtain Tolman's law by considering a fluid in rest with 
$U^\mu = \left({c}/{\sqrt{-g_{00}}}, \vec{0}\right)$.
Indeed, by taking into account the existence of a timelike Killing vector amounts to a stationary metric, the acceleration equation reduces to
\begin{eqnarray}
\dot{U}^\mu = -\frac{c^2}{g_{00}}\Gamma_{00}^\mu = \frac{c^2}{2g_{00}}g^{\mu\nu}g_{00,\nu}.
\label{eq117}
\end{eqnarray}
Hence, from the right-hand expression of (\ref{eq112}) and from (\ref{eq117}) we get
\begin{gather}
c^2 g^{\mu\nu}[(\ln\sqrt{-g_{00}}T)_{,\mu}]=0,
\label{eq121}
\end{gather}
and it follows Tolman's law
\begin{gather}
\sqrt{-g_{00}}T = {\rm constant}.
\label{eq122a}
\end{gather}

The equilibrium condition given by the first equation (\ref{eq106}) together with Tolman's law (\ref{eq122a}) implies Klein's law \cite{klein}, namely $\sqrt{-g_{00}}\mu = {\rm constant}$. 

\section{Equations of motion for a particle in Reissner-Nordstr\"om metric}
\label{sec02}

The Reissner-Nordstr\"om metric reads
\begin{gather}
ds^2 = - \left(1-\frac{2M'}{r}+\frac{Q^2}{r^2}\right)(dx^0)^2 + \frac{1}{\left(1-\frac{2M'}{r}+ \frac{Q^2}{r^2}\right)}dr^2 \nonumber\\
+ r^2(d \theta^2 + \sin^2 \theta d \phi^2) = -c^2 d\tau^2,
\label{eq125}
\end{gather}
where $M' = GM/c^2$ and $Q^2 = {q^2 G}/{(4\pi \epsilon_0 c^4)}$.  $M$ and $q$ denote the mass and the electric charge of a massive and charged object, $G $ is the gravitational constant, $c$ the speed of light in vacuum and $\epsilon_0 $ the vacuum permittivity. The coordinate $x^0$ corresponds to the temporal coordinate, and $(r, \theta, \phi)$ are the spatial coordinates in the spherical coordinate system. The invariant $ds^2 =-c^2 d\tau^2$ defines the proper time $\tau$.

The event horizon for the Reissner-Norstr\"om metric is defined by the expression \cite{chandrasekhar}:
\begin{equation}
\Delta = r^2 - 2M'r + Q^2.
\label{eq45}
\end{equation}
The roots of (\ref{eq45}) are $r_+ = M'+\sqrt{M'^2-Q^2}$ and $r_- = M'-\sqrt{M'^2-Q^2}$. These roots will be real and distinct if $M'^2 > Q^2$. In this work we will adopt as the Reissner-Norstr\"om radius $R_{RN}=M'+\sqrt{M'^2-Q^2}$, i.e., the root $r_+$ of (\ref{eq45}). This choice was taken in order to avoid the interchange of the spacelike and timelike behavior of the coordinates $r$ and $t$, that occurs in the region between $r_+$ and $r_-$ \cite{dinverno}. Other choices are possible but it's needed to deal with points where the behavior of spacelike and timelike coordinates change.

The orbital motion of a test particle with rest mass $m$ in the plane $\theta = \pi/2$ around a source of gravitational field with rest mass $M$ and electric charge $q$ is described by the Lagrangian
\begin{gather}
\mathcal{L} = \frac{m}{2} \left[ \frac{1}{\left(1-\frac{2M'}{r}+ \frac{Q^2}{r^2}\right)} \left( \frac{dr}{d\tau}\right)^2 + r^2 \left(\frac{d\phi}{d\tau}\right)^2 \right. \nonumber\\
\left. - \left(1-\frac{2M'}{r}+ \frac{Q^2}{r^2}\right) \left(\frac{dx^0}{d\tau}\right)^2 \right],
\label{eq126}
\end{gather}
while the generalized momenta corresponding to the cyclic coordinates $x^0$ and $\phi$ are expressed by
\ben
p_0 = \frac{\partial \mathcal{L}}{\partial \left(\frac{dx^0}{d\tau}\right)} = -m\left(1-\frac{2M'}{r}+\frac{Q^2}{r^2}\right)\left(\frac{dx^0}{d\tau}\right) = -\frac{E}{c},
\label{eq127}
\\
p_\phi = \frac{\partial \mathcal{L}}{\partial \left(\frac{d\phi}{d\tau}\right)} = mr^2\left(\frac{d\phi}{d\tau}\right) = l_\phi.
\label{eq128}
\een
In the above equations $E$ is the energy of the particle and $l_\phi$  its angular momentum. In order to keep the dimensions, we will keep the value of $c$ in our calculations. The center of the spherical coordinate system are defined as the center of the source of the gravitational field.

By introducing  the dimensionless energy $\epsilon = E/mc^2$ and the dimensionless angular momentum $l = l_\phi/mc$ and taking into account a circular orbit of constant radius, one can find  from (\ref{eq125}), (\ref{eq127}) and (\ref{eq128}) the relation:
\begin{equation}
\epsilon^2 = \left(1-\frac{2M'}{r}+\frac{Q^2}{r^2}\right)\left(1 + \frac{l^2}{r^2} \right).
\label{eq129}
\end{equation}
Furthermore, from the equation of motion for the test particle it follows that
\begin{equation}
\left(\frac{dr}{dc\tau}\right)^2 + V = \epsilon^2,
\label{eq130}
\end{equation}
where $V$ is an effective potential defined by
\begin{equation}
V = \left(1-\frac{2M'}{r}+\frac{Q^2}{r^2}\right)\left(1 + \frac{l^2}{r^2} \right).
\label{eq131}
\end{equation}

From the extreme values of the effective potential (maximum and minimum points of the function, where $dV/dr=0$) one can get the possible circular orbits for the test particle, namely
\begin{equation}
l^2 = \frac{r^2 \left(M'r-Q^2\right)}{r^2 - 3M'r+2Q^2}.
\label{eq132}
\end{equation}

Now the insertion of   (\ref{eq132}) into (\ref{eq129}) leads to  the following expression for the dimensionless energy:
\begin{equation}
\epsilon^2 = \left(1-\frac{2M'}{r}+\frac{Q^2}{r^2}\right)^2\left(\frac{r^2}{r^2 - 3M'r+2Q^2} \right).
\label{eq133}
\end{equation}
Equations (\ref{eq132}) and (\ref{eq133}) correspond to the values of the angular momentum and energy of the test particle with rest mass $m$ in orbital motion with $\theta = \pi/2$.

By combining  the expressions (\ref{eq128}) and (\ref{eq132}) for $d\phi/d\tau$  and  for the angular momentum $l$  and   integrating  the resulting equation yields
\begin{equation}
\phi = \frac{1}{r}\sqrt{\frac{M'r-Q^2}{r^2-3M'r+2Q^2}}c\tau,
\label{eq134}
\end{equation}
where the integration constant disappear with a simple redefinition of the variable $\phi$. The angular frequency $\omega_\phi$ for the particle motion can be defined as:
\begin{equation}
\omega_\phi \equiv \frac{d \phi}{d \tau} = \omega_N \sqrt{\frac{r\left(r-Q^2/M'\right)}{r^2-3M'r+2Q^2}}, \qquad \omega_N = \sqrt{\frac{GM}{r^3}}.
\label{eq135}
\end{equation}
In (\ref{eq135}), $\omega_N$ is the Newtonian frequency in the limits $r\gg M'$ and $r\gg Q$. Note that  the definition for $M'=GM/c^2$ was used.

According to ref. \cite{wald} another oscillation frequency can be obtained when the particle is slightly displaced from the circular motion in the radial direction. This radial frequency $\omega_r$ is defined in terms of the second derivative of the effective potential, namely
\begin{equation}
\omega_r \equiv \sqrt{\frac{c^2}{2}\frac{d^2 V}{dr^2}} = \omega_N \sqrt{\frac{r^2-6M'r+9Q^2-4Q^4/M'r}{r^2-3M'r+2Q^2}}.
\label{eq137}
\end{equation}
Here we have used  the expression (\ref{eq132}) for $l^2$.

By considering a vanishing electric charge $Q=0$ the two frequencies (\ref{eq135}) and (\ref{eq137}) reduce to
the expressions   for a Schwarzschild metric \cite{wald}:
\ben
\omega_\phi = \omega_N \sqrt{\frac{r}{r-3M'}},\qquad
\omega_r =   \omega_N \sqrt{\frac{r-6M'}{r-3M'}}.
\een

As was pointed out by Wald \cite{wald} the difference of the two frequencies implies in a precession of the particle motion inside the range of the orbital plane. In the Reissner-Nordstr\"om metric the precession  frequency $\omega_p$ is given by
\ben
\omega_p=\omega_\phi\left(1-\frac{\omega_r}{\omega_\phi}\right)=\omega_\phi\left(1-\sqrt{\frac{1-6M'/r+9Q^2/r^2-4Q^4/M'r^3}{1-Q^2/M'r}}\right).
\een

There are other frequencies that can be calculated from the Lagrangian of the free particle (for example, frequencies related to the coordinate time $x^0 = t$ such as $d \phi / d t$). In the present work we calculated $\omega_r$ and $\omega_\phi$ in order to compare them with the frequencies obtained in the next section.

\section{Gas in circular motion with geodesic deviation in Reissner-Nordstr\"om space-time}
\label{sec02a}

We follow ref. \cite{zimdahl15} and consider a rarefied gas inside a spacecraft which is in a circular orbit around an object with mass $M$ and electric charge $q$. We assume that the contributions of the spacecraft and the gas to the gravitational field can be neglected and consider that the center of mass of the gas is moving on a circular geodesic of the Reissner-Nordstr\"om metric. The geometry and symmetries of the problem are explored by using the Fermi normal coordinates (exploring the geodesic motion) and spherical coordinates (exploring the circular orbit).

\subsection{Fermi Normal Coordinates}

In 1922, Enrico Fermi showed that, given any curve in a Riemannian manifold, it's possible to introduce coordinates near this curve in such a way that the Christoffel symbols vanish along the curve, leaving the metric locally rectangular \cite{manasse63}. If we consider the curve in question a geodesic, this particularization of Fermi's idea leads to a coordinate system known as Fermi normal coordinates.

In order to describe the local gravitational effects in the vicinity of the geodesic, a smart choice for an observer at the center of mass is to use Fermi normal coordinates \cite{manasse63}, which are comoving and time-orthogonal coordinates with the center of mass at rest in the origin.  The proper time $\tau$ of the center on the geodesic is the time coordinate while the spatial coordinates are orthogonal space-like geodesics parametrized by the proper distance.

We begin by writing the non-vanishing components of the Riemann tensor for Reissner-Nordstr\"om metric in the case where  $\theta=\pi/2$:
\begin{gather}
R_{\bar0\bar1\bar0\bar1} = -\frac{2M'}{r^3}+\frac{3Q^2}{r^4}, \quad R_{\bar2\bar3\bar2\bar3} = 2M'r-Q^2, \nonumber\\
R_{\bar0\bar2\bar0\bar2} = R_{\bar0\bar3\bar0\bar3} = \frac{M'-Q^2/r}{r}\left(1-\frac{2M'}{r}+\frac{Q^2}{r^2}\right),\nonumber\\
R_{\bar1\bar2\bar1\bar2} = R_{\bar1\bar3\bar1\bar3} = -\frac{M'-Q^2/r}{r}\left(1-\frac{2M'}{r}+\frac{Q^2}{r^2}\right)^{-1}.
\label{eq138}
\end{gather}

In the equations (\ref{eq138}), the overbar denotes the original Reissner-Nordstr\"om coordinates according to (\ref{eq125}). The numerical index $0$ denotes the temporal coordinate, and the indexes $(1,2,3)$ denote spatial coordinates.

For a circular geodesic in a Reissner-Nordstr\"om field, the Fermi normal tetrads has the same form of the tetrads for a Schwarzschild field \cite{collas07, parker82}, and are given by:
\ben
(e^{\bar\alpha}_{\hat{0}}) = \left(\frac{\epsilon}{X},0,0, \frac{l}{r^2}\right), \\
(e^{\bar\alpha}_{\hat{1}}) =\left( -\frac{l \sin(\alpha \phi)}{r \sqrt{X}}, \sqrt{X} \cos(\alpha \phi), 0, - \frac{\epsilon \sin(\alpha \phi)}{r \sqrt{X}} \right),\\
(e^{\bar\alpha}_{\hat{2}}) = \left(0,0,\frac{1}{r},0\right), \\
(e^{\bar\alpha}_{\hat{3}}) =\left(\frac{l \cos(\alpha \phi)}{r \sqrt{X}}, \sqrt{X} \sin(\alpha \phi), 0, \frac{\epsilon \cos(\alpha \phi)}{r \sqrt{X}} \right),
\label{eq139}
\een
where the following abbreviations  were introduced:

\begin{equation}
\alpha = \frac{\sqrt{r^2 - 3M'r+2Q^2}}{r}, \quad X = 1-\frac{2M'}{r}+\frac{Q^2}{r^2}.
\label{eq140}
\end{equation}

Given some initial time, $(e^{\bar\alpha}_{\hat{1}})$ points to the radial direction and 
$(e^{\bar\alpha}_{\hat{3}})$ points to the tangential direction, while $(e^{\bar\alpha}_{\hat{2}})$ is always perpendicular to the orbital plane. On the circular geodesics

\begin{equation}
g_{\alpha \beta}\frac{\partial x^\alpha}{\partial x^{\hat{\mu}}} \frac{\partial x^\beta}{\partial x^{\hat{\nu}}} = \eta_{\hat{\mu} \hat{\nu}}
\label{eq141}
\end{equation}

\noindent is valid, where $\eta_{\hat{\mu} \hat{\nu}}$ is the Minkowski metric. The tetrads are parallel transported along the circular geodesic

\begin{equation}
\frac{D e^{\bar\mu}_{\hat{\alpha}}}{d \tau} = 0,
\label{eq141a}
\end{equation}

\noindent where the operator $D /{d\tau}$ is the absolute derivative with respect to the proper time $\tau$ \cite{dinverno}.

The components of Riemann tensor in Fermi normal coordinates can be calculated by using the expression \cite{manasse63}:
\begin{equation}
R_{\hat{\mu} \hat{\nu}\hat{\sigma}\hat{\tau}} = R_{\bar\alpha\bar \beta\bar \gamma \bar\delta}\hat{e}_{\hat{\mu}}^{\bar\alpha} \hat{e}_{\hat{\nu}}^{\bar\beta} \hat{e}_{\hat{\sigma}}^{\bar\gamma} \hat{e}_{\hat{\tau}}^{\bar\delta},
\label{eq142}
\end{equation}
where the hat over the indexes refers to coordinates in the Fermi system. The non-vanishing components of the Riemann tensor are:
\ben
&&R_{\hat{0} \hat{1}\hat{0}\hat{1}} = \frac{\left(2r^2Q^2-M'rQ^2-r^3M'\right)}{2r^4 \left(r^2-3M'r+2Q^2\right)} \nonumber
\\
&&\qquad+\frac{\cos\left(2\alpha\phi\right)}{2r^4 \left(r^2-3M'r+2Q^2\right)}\left(4Q^2-3M'r\right)\left(r^2-2M'r+Q^2\right),\\
&&R_{\hat{0} \hat{2}\hat{0}\hat{2}}=\frac{M'r-Q^2}{r^2\left(r^2-3M'r+2Q^2\right)},\\
&&R_{\hat{0} \hat{3}\hat{0}\hat{3}} = \frac{\left(2r^2Q^2-M'rQ^2-r^3M'\right)}{2r^4 \left(r^2-3M'r+2Q^2\right)} \nonumber
\\
&&\qquad-\frac{\cos\left(2\alpha\phi\right)}{2r^4 \left(r^2-3M'r+2Q^2\right)}\left(4Q^2-3M'r\right)\left(r^2-2M'r+Q^2\right),\\
&&R_{\hat{0} \hat{1}\hat{0}\hat{3}} = \frac{\left(4Q^2-3M'r\right)\left(r^2-2M'r+Q^2\right)\sin \left(2\alpha\phi\right)}{2r^4 \left(r^2-3M'r+2Q^2\right)},\\
&&R_{\hat{0} \hat{1}\hat{1}\hat{3}} = \frac{\sqrt{\left(r^2-2M'r+Q^2\right)\left(M'r-Q^2\right)}}{r^4\left(r^2-3M'r+2Q^2\right)} \left(3M'r-4Q^2\right)\cos(\alpha\phi),\\
&&R_{\hat{0} \hat{2}\hat{1}\hat{2}} = -\frac{\sqrt{\left(r^2-2M'r+Q^2\right)\left(M'r-Q^2\right)}}{r^4\left(r^2-3M'r+2Q^2\right)} \left(3M'r-2Q^2\right)\sin(\alpha\phi).
\label{eq143}
\een
The following relations also hold:
\begin{gather}
R_{\hat{0}\hat{3}\hat{1}\hat{3}} = -\frac{(4Q^2-3M'r)}{(2Q^2-3M'r)} R_{\hat{0} \hat{2}\hat{1}\hat{2}},\quad R_{\hat{0}\hat{2}\hat{2}\hat{3}} = -\frac{(2Q^2-3M'r)}{(4Q^2-3M'r)} R_{\hat{0} \hat{1}\hat{1}\hat{3}}  ,\\
R_{\hat{1}\hat{2}\hat{2}\hat{3}} = -\frac{(2Q^2-3M'r)}{(4Q^2-3M'r)} R_{\hat{0} \hat{1}\hat{0}\hat{3}},\quad R_{\hat{1}\hat{3}\hat{1}\hat{3}} = \frac{(2Q^2-r^2)}{r^2} R_{\hat{0} \hat{2}\hat{0}\hat{2}}  ,\\
R_{\hat{2}\hat{3}\hat{2}\hat{3}} = \frac{2Q^2(r^2-2M'r+Q^2)\cos^2(\alpha \phi)}{r^4(r^2-3M'r+2Q^2)} - R_{\hat{0} \hat{1}\hat{0}\hat{1}},\\
R_{\hat{1}\hat{2}\hat{1}\hat{2}} = \frac{2Q^2(r^2-2M'r+Q^2)\sin^2(\alpha \phi)}{r^4(r^2-3M'r+2Q^2)} - R_{\hat{0} \hat{3}\hat{0}\hat{3}}.
\label{eq143b}
\end{gather}

The component $g_{\hat{0}\hat{0}}$ of the metric tensor up second order in geodesic deviation in the Fermi normal coordinates is given by (see refs. \cite{misner,manasse63}):
\begin{equation}
g_{\hat{0}\hat{0}} = -1 - R_{\hat{0}\hat{n}\hat{0}\hat{m}}x^{\hat{n}} x^{\hat{m}}.
\label{eq144}
\end{equation}

Because of the non-vanishing component $R_{\hat{0} \hat{1}\hat{0}\hat{3}}$, the second order contribution in $g_{\hat{0}\hat{0}}$ is not diagonal. Here we use the same procedure adopted in ref. \cite{zimdahl15} to obtain a simpler expression by  performing a tetrad rotation around the direction $x^{\hat{2}}$ perpendicular to the orbital plane. The transformation is given by
\ben\label{eq145a}
{\bf E}_{\bar{0}} \equiv {\bf e}_{\hat{0}}, \qquad
{\bf E}_{\bar{1}} \equiv {\bf e}_{\hat{1}} \cos(\alpha\phi) + {\bf e}_{\hat{3}} \sin(\alpha\phi),\\
{\bf E}_{\bar{2}} \equiv {\bf e}_{\hat{2}},\qquad
{\bf E}_{\bar{3}} \equiv {\bf e}_{\hat{1}} \sin(\alpha\phi) - {\bf e}_{\hat{3}} \cos(\alpha\phi),
\label{eq145b}
\een
where $(E^{\bar\alpha}_{\hat{1}})$ always shows in the radial direction and $(E^{\bar\alpha}_{\hat{3}})$ always shows in the tangential direction. So that the components of the Riemann tensor in the rotated system can be  calculated through

\begin{equation}
R_{abcd} = \frac{\partial {\bf E}_{\bar{\mu}}}{\partial {\bf e}_{\hat{a}}} \frac{\partial {\bf E}_{\bar{\nu}}}{\partial {\bf e}_{\hat{b}}} \frac{\partial {\bf E}_{\bar{\sigma}}}{\partial {\bf e}_{\hat{c}}} \frac{\partial {\bf E}_{\bar{\tau}}}{\partial {\bf e}_{\hat{d}}} R_{\hat{\mu} \hat{\nu}\hat{\sigma}\hat{\tau}}.
\label{eq146}
\end{equation}

In this frame, the relevant Riemann tensor components for the calculation of $g_{00}$ take a simpler form:
\ben\label{eq147a}
R_{0101} = \frac{2Q^4 - 6M'rQ^2 + 3Q^2r^2 - 2M'r^3 + 3M'^2r^2}{r^4(r^2-3M'r+2Q^2)},\\
R_{0202} = \frac{M'r-Q^2}{r^2(r^2-3M'r+2Q^2)},\qquad
R_{0303} = \frac{M'r-Q^2}{r^4}.
\label{eq147b}
\een

Using the above expressions for the components of the Riemann tensor in (\ref{eq144}) we obtain:
\begin{gather}
g_{00} = -1 + \frac{2M'r^3 - 3(M'^2+Q^2)r^2+6M'Q^2r-2Q^4}{r^4(r^2-3M'r+2Q^2)}(x^1)^2\nonumber\\
-\frac{M'r-Q^2}{r^2(r^2-3M'r+2Q^2)}(x^2)^2 - \frac{M'r-Q^2}{r^4}(x^3)^2.
\label{eq148}
\end{gather}

One important consequence of the choice of the tetrads (\ref{eq145a}) and (\ref{eq145b}) is that they are no longer parallel transported. From the condition of parallel transport of the tetrads ${D {\bf e}_{\hat{1}}}/{d\tau}=0$ and ${D {\bf e}_{\hat{3}}}/{d\tau}=0$
and the relationships (\ref{eq145a}) and (\ref{eq145b}) it follows that
\ben
\frac{D {\bf E}_{\bar{1}}}{d\tau} = \sqrt{\frac{M'-Q^2/r}{r^3}} {\bf E}_{\bar{3}}, \quad \frac{D {\bf E}_{\bar{3}}}{d\tau} = -\sqrt{\frac{M'-Q^2/r}{r^3}} {\bf E}_{\bar{1}}.
\een
The above equations imply that the tetrad transformations describe a rotation with frequency
\ben
\omega=\sqrt{\frac{M'-Q^2/r}{r^3}},
\een
with non-vanishing Christoffel symbols on the geodesic and mixed space-time terms linear in $x^i$ in the metric, namely (see ref. \cite{zimdahl15})
\ben\label{a1}
\Gamma_{03}^1=\sqrt{\frac{M'-Q^2/r}{r^3}}\qquad \Gamma_{01}^3=-\sqrt{\frac{M'-Q^2/r}{r^3}},
\\\label{a2}
g_{01}=\sqrt{\frac{M'-Q^2/r}{r^3}}x^3\qquad
g_{03}=-\sqrt{\frac{M'-Q^2/r}{r^3}}x^1.
\een

\subsection{Geodesic deviation}
\label{sec02b}

A geodesic fluid motion is characterized by the condition that the four-velocity obeys the equation $\dot U^\mu=0$. The  four-velocity of a gas at equilibrium is not compatible with the equation of the geodesic motion, since it depends on the gradient of the temperature (see ref. \cite{zimdahl15,cercignani}), namely
\ben
\dot U_\mu+\frac{c^2}T\nabla_\mu T=0.
\een

However, that terms that "perturb" the geodesic behavior are linear in distance, so the equation of geodesic deviation can be applied in our problem \cite{zimdahl15}. The general  equation for geodesic deviation for a vector  $\xi^\alpha$ orthogonal to the geodesics is
\begin{equation}
\frac{D^2 \xi^\alpha}{d\tau^2}+ {R^\alpha}_{\gamma \mu \nu}U^\gamma U^\nu \xi^\mu = 0,
\label{eq150}
\end{equation}
where ${D^2 \xi^\alpha}/{d\tau^2}$ is given by
\ben\nonumber
\frac{D^2\xi^\alpha}{d\tau^2}=\frac{d^2\xi^\alpha}{d\tau^2}+\Gamma^\alpha_{\beta\gamma,\rho}\,
\xi^\beta\frac{dx^\rho}{d\tau}\frac{dx^\gamma}{d\tau}
+2\Gamma^\alpha_{\beta\gamma}\frac{d\xi^\beta}{d\tau}\frac{dx^\gamma}{d\tau}
\\\label{n1}
+\Gamma^\alpha_{\beta\gamma}\Gamma^\beta_{\rho\sigma}\xi^\rho\frac{dx^\sigma}{d\tau}
\frac{dx^\gamma}{d\tau}+{\Gamma^\alpha_{\beta\gamma}\xi^\beta\frac{d^2x^\gamma}{d\tau^2}}.
\een

The last term in (\ref{n1}) can be written as
\ben
\Gamma^\alpha_{\beta\gamma}\xi^\beta\frac{d^2x^\gamma}{d\tau^2} = - \Gamma^\alpha_{\beta\gamma}\xi^\beta
\left[\Gamma^{\gamma}_{\kappa\lambda}U^{\kappa}U^{\lambda} + \frac{c^{2}}{T}\nabla^{\gamma}T\right],
\een
thanks to the second equilibrium equation (\ref{eq112}) which can be rewritten as
\ben
\dot U^\gamma = \frac{DU^\gamma}{d\tau} = \frac{D^2x^\gamma}{d\tau^2}=- \frac{c^{2}}{T}\nabla^{\gamma}T,\qquad
\frac{d^2x^\gamma}{d\tau^2} = - \Gamma^{\gamma}_{\kappa\lambda}U^{\kappa}U^{\lambda} - \frac{c^{2}}{T}\nabla^{\gamma}T.
\een

By neglecting the second-order contributions of the temperature gradient, approximating the four-vector  by $(U^\mu) = (c,\vec{0})$  and neglecting the second-order contributions to $g_{00}=-1+\mathcal{O}(x^2)$, the expression (\ref{eq150}) can be written as
\ben
\frac{d^2x^a}{d\tau^2}+2c\Gamma_{\beta 0}^a\frac{dx^\beta}{d\tau}-c^2\Gamma_{\beta \gamma}^a x^\beta\Gamma_{00}^\gamma + c^2\Gamma_{\beta 0}^a \Gamma_{\rho 0}^\beta x^\rho +c^2g^{a\beta}R_{\beta 0 \mu 0}x^\mu = 0,
\label{eq158}
\een
when one considers only the perturbation terms  with linear dependence in the distance and makes use of the expression 
\ben\label{n2}
{R^\alpha}_{\gamma\mu\nu}U^\gamma U^\nu \xi^\mu=c^2g^{\alpha\beta}R_{\beta0\mu0}\xi^\mu.
\een

In (\ref{eq158}) the spatial components of $\xi^a$ were identified with the components of the tetrad system $x^a$.

Now by using the expressions for the components of the Riemann tensor (\ref{eq147a}), (\ref{eq147b}), Christoffel symbol (\ref{a1}) and metric tensor (\ref{a2}) and replacing $a = 1,2,3$ in (\ref{eq158}), we get a system of differential equations for the spatial components of $x^a$:
\ben
&&\frac{d^2 x^1}{d\tau^2}+c^2\left[\frac{4Q^4-11M'rQ^2+4Q^2r^2-3M'r^3+6M'^2r^2}{r^4\left(r^2-3M'r+2Q^2\right)}
\right]x^1  \nonumber
\\\label{eq161a}
&&\qquad -2c\sqrt{\frac{M'-Q^2}{r^4}}\frac{dx^3}{d\tau} =0,\\\label{eq161b}
&&\frac{d^2x^2}{d\tau^2}+ \frac{c^2(M'r-Q^2)}{r^2(r^2-3M'r+2Q^2)}x^2 = 0,\\
&&\frac{d^2 x^3}{d\tau^2} + 2c\sqrt{\frac{M'r-Q^2}{r^4}}\frac{dx^1}{d\tau}=0.
\label{eq161c}
\een

From the above system of differential equations we infer that  (\ref{eq161b}) decouples from the two other equations and has a  real solution given by
\ben\label{s1}
x^2 = x_0^2 \sin(\Omega\tau),
\een
where $\Omega$ denotes the frequency of the harmonic motion of the $x^2$ component
\ben\label{s2}
\Omega=\Omega_N\sqrt{\frac{r(r-Q^2/M')}{r^2-3M'r+2Q^2}}, \qquad \Omega_N = \sqrt{\frac{GM}{r^3}}.
\een
Note that $\Omega$ coincides with the orbital frequency $\omega_\phi$ for the test particle in (\ref{eq135}).

The real solutions of the coupled system of equations (\ref{eq161a}) and (\ref{eq161c}) read
\ben\label{s3a}
&&x^1 = x_0^1 \sin(\omega\tau),\\\label{s3b}
&&x^3 = 2\sqrt{\frac{(M'r-Q^2)(r^2-3M'r+2Q^2)}{M'r^3 - 6M'^2r^2+9Q^2M'r-4Q^2}}x_0^1 \cos(\omega\tau).
\een
Here $\omega$ characterizes oscillations in the plane $(x^1, x^3)$
\ben\label{s4}
\omega = \omega_N \sqrt{\frac{r^3-6M'r^2+9Q^2 r-4Q^4/M'}{r(r^2-3M'r+2Q^2)}}, \qquad \omega_N = \sqrt{\frac{GM}{r^3}},
\een
and it coincides with the radial frequency $\omega_r$ for the test particle (\ref{eq137}). This frequency refers to oscillations in the tangential direction and describes an ellipse in the $(x^1, x^3)$ plane. However here there is no precession in the orbital plane for a comoving observer.

In the limit $r\gg M'$ and $r\gg Q$ these frequencies coincide, i.e., $\Omega=\Omega_N=\omega=\omega_N$.

\subsection{Frequencies and orbit stability}

By analyzing the oscillation frequencies $\Omega$ and $\omega$ given by  (\ref{s2}) and (\ref{s4}), we can infer that the roots of the following polynomials:
\ben\label{eq170}
r^3-6M'r^2+9Q^2 r-4Q^4/M'=0,
\\\label{eq171}
r^2-3M'r+2Q^2=0,
\een
represent critical points for the analysis of the frequencies, because these polynomials are the denominators of frequency expressions (\ref{s2}) and (\ref{s4}). The polynomial in equation (\ref{eq170}) has one real root, namely
\begin{equation}
r_{\omega_0} = 2M'+ \frac{4M^{\prime2}-3Q^2}{D^{1/3}} + D^{1/3},
\label{eq176}
\end{equation}
where
\begin{equation}
D = 8M^{\prime3}-9M'Q^2+2Q^4/M'+ \sqrt{5M'^2 Q^4-9Q^6+4Q^8/M'^2}.
\label{eq177}
\end{equation}

The polynomial in equation (\ref{eq171}) has two real roots:
\begin{equation}
r_{\omega_-} = \frac{1}{2}(3M'-\sqrt{9M'^2-8Q^2}), \quad r_{\omega_+} = \frac{1}{2}(3M'+\sqrt{9M'^2-8Q^2})
\label{eq178}
\end{equation}

It's important to note that the radii from expressions (\ref{eq176}) and (\ref{eq178}) are not related with the event horizons defined by the roots of equation (\ref{eq45}), but represent limit points of orbit instability. Another important point for this analysis is when $r=Q^2/M'$, since at this point  $\Omega = 0$. At this point we can analyze the following regimes for orbital stability:

\begin{itemize}
	\item {$r > r_{\omega_0}$}: in this region, all circular orbits are stable. When $r$ reaches the limit $r \to  r_{\omega_0}$, we have $\omega \to 0$ and the oscillations are frozen in the $x^1-x^3$ plane, being restricted to the $x^2$ plane.
	\item {$r_{\omega_+} < r < r_{\omega_0} $}: in this region, there exist unstable circular trajectories. Here, $\omega$ becomes imaginary and we have exponential instabilities in the $x^1-x^3$ plane. The oscillations grows to infinity when the limit $r = r_{\omega_+}$ is approached.
	\item {$r_{\omega_-} < r < r_{\omega_+} $}: in this region there also unstable trajectories. But here, $\Omega$ becomes imaginary and the exponential instabilities are in the $x^2$ plane. When $r$ reaches the limit $r \to  Q^2/M'$, we have $\Omega \to 0$ and the oscillations are frozen in the $x^2$ plane, being restricted to the $x^1-x^3$ plane.
\end{itemize}

If the charge values $Q$ are small, we can expand the frequencies $\omega$ and $\Omega$ in series around $Q=0$, yielding
\ben
\frac{\omega}{\omega_N} = \sqrt{\frac{r-6M'}{r-3M'}} + \frac{Q^2(7r-15M')}{2r \sqrt{(r-6M')(r-3M')^3}} + \mathcal{O}(Q^3),
\label{eq179}
\\
\frac{\Omega}{\Omega_N} = \sqrt{\frac{r}{r-3M'}} - \frac{Q^2(r-M')}{2M' \sqrt{r(r-3M')^3}} + \mathcal{O}(Q^3).
\label{eq180}
\een

In the limit $Q \to 0$  the frequencies $\Omega$ and $\omega$ reduce to those found in ref. \cite{zimdahl15} for a  Schwarzschild metric.

\section{Temperature oscillations}
\label{sec03}

Let us turn to the problem of a gas at equilibrium inside a spacecraft in a circular geodesic motion in a spacetime described by the Reissner-Nordstr\"om metric. As was pointed out the temperature field obeys Tolman's law $\sqrt{-g_{00}}T = {\rm constant}$ and if $T_0$ is a constant equilibrium temperature on the geodesics, we can approximate the temperature field in the vicinity of the central geodesic by using  (\ref{eq148}), yielding
\begin{gather}
\frac{T}{T_0} \approx 1 + \frac{2M'r^3 - 3(M'^2+Q^2)r^2+6M'Q^2r-2Q^4}{2r^4(r^2-3M'r+2Q^2)}(x^1)^2\nonumber\\
-\frac{M'r-Q^2}{2r^2(r^2-3M'r+2Q^2)}(x^2)^2 - \frac{M'r-Q^2}{2r^4}(x^3)^2.
\label{eq149}
\end{gather}

Now the  replacement of the values of $x^1$, $x^2$ and $x^3$ from  (\ref{s3a}), (\ref{s1}) and (\ref{s3b}) in the temperature profile (\ref{eq149}) results:
\ben
\frac{T-T_0}{T_0} = \Delta(\tau)= \frac{M'(x_0^1)^2}{4r^3}\left[(A-B) - (A+B)\cos(2\omega\tau)\right.\nonumber\\ \left. -C\left(\frac{x_0^2}{x_0^1}\right)^2(1-\cos(2\Omega\tau)) \right],
\label{eq167}
\een
where the coefficients  $A$, $B$ and $C$ are  given by:
\ben
A = \frac{2M'r^3 - 3(Q^2+M'^2)r^2 + 6M'Q^2r - 2Q^4}{M'r(r^2-3M'r+2Q^2)}\\
B = \frac{4(M'r-Q^2)^2(r^2-3M'r+2Q^2)}{M'r(M'r^3-6M'^2r^2+9Q^2M'r-4Q^4)}\\
C = \frac{r(M'r-Q^2)}{M'(r^2-3M'r+2Q^2)}.
\label{eq169}
\een
The coefficients $A$, $B$ and $C$  as well as the frequencies $\Omega$ and $\omega$,  can be determined from knowledge of the mass, electric charge and radius of the gravitational field source. The constants $x_0^1$ e $x_0^2$ are free parameters.

Note that $\Delta(\tau)$ is a dimensionless and normalized amplitude of the temperature oscillations. It was chosen because it can be calculated even if we do not know the equilibrium temperature. In the next subsections this amplitude will be analyzed for some theoretical models from the literature that consider the existence of electric charge in some compact objects. The problem consists of a gas in circular orbit around a charged massive object that produces a strong gravitational field. To avoid the region between the event horizons of the Reissner-Nordstr\"om metric, where occurs the interchange of the spacelike and timelike components of the metric tensor \cite{dinverno}, circular orbits with $r=5R_{RN}$, where $R_{RN}$ is the Reissner-Nordstr\"om radius, are considered. The choice of this multiple of the Reissner-Nordstr\"om radius is arbitrary but also takes into account the orbit stability. We also take $x_0^1 = x_0^2 = 1$.

For each of the analyzed models, graphs are plotted relating the normalized amplitude $\Delta(\tau)$ and the proper time $\tau$, using different mass and electric charge configurations. For each case the ratio between the frequencies $\Omega/\omega$ is also calculated.

\subsection{Charged compact stars and  charged quark stars}
\label{sec03a}

In 2003 Ray et al \cite{ray03} presented a model to describe the effect of electric charge on compact stars, assuming that the charge distribution is proportional to the mass density. This model consider a polytropic equation of state for charged stars. Based on the mass and electric charge values for compact stars provided from ref. \cite{ray03} the table \ref{tab01} was elaborated, containing, in addition to mass and electric charge, the orbit radius $r=5R_{RN}$ and the ratio between frequencies $\Omega/\omega$. 

\begin{table}
	\caption{Charged compact stars.}
	{\begin{tabular}{cccccc} \
			$M(M_{\odot})$ & $r = 5 R_{RN}(km)$ & $q(\times 10^{20} C)$ & $\omega(kHz)$ & $\Omega(kHz)$ &$\Omega/\omega$\\ 
			$1.430$&$21.030$&$0.259$&$3.413$&$5.399$&$1.582$\\
			
			$1.765$&$24.275$&$1.517$&$3.027$&$4.826$&$1.595$\\
			
			$2.728$&$33.769$&$3.434$&$2.250$&$3.657$&$1.626$\\
			
			$5.248$&$59.560$&$7.576$&$1.306$&$2.173$&$1.663$\\
			
			$12.150$&$132.240$&$18.314$&$0.594$&$1.002$&$1.686$\\ 
		\end{tabular} \label{tab01}}
\end{table}

In 2009 Negreiros et al \cite{negreiros09} presented a model for electrically charged quark stars. These stars are formed from a compression process similar to that forming neutron stars, but much more intense, where the neutrons decay to the quarks that constitute them.
Based on the values provided by ref. \cite{negreiros09}, the table \ref{tab02} was elaborated, containing values of mass, electric charge, orbit radius $r=5R_{RN}$ and ratio between the frequencies $\Omega/\omega$ for quark stars.

\begin{table}
	\caption{Charged quark stars.}
	{\begin{tabular}{cccccc} 
			$M(M_{\odot})$ & $r = 5 R_{RN}(km)$ & $q(\times 10^{20} C)$ & $\omega(kHz)$ & $\Omega(kHz)$& $\Omega/\omega$\\ 
			$2.02$&$29.791$&$0$&$2.407$&$3.806$&$1.581$\\
			
			$2.07$&$29.923$&$0.989$&$2.414$&$3.824$&$1.584$\\
			
			$2.15$&$30.362$&$1.486$&$2.398$&$3.810$&$1.589$\\
			
			$2.25$&$30.824$&$1.982$&$2.387$&$3.808$&$1.595$\\ 
		\end{tabular} \label{tab02}}
\end{table}

\begin{figure}
	\includegraphics[width=6cm]{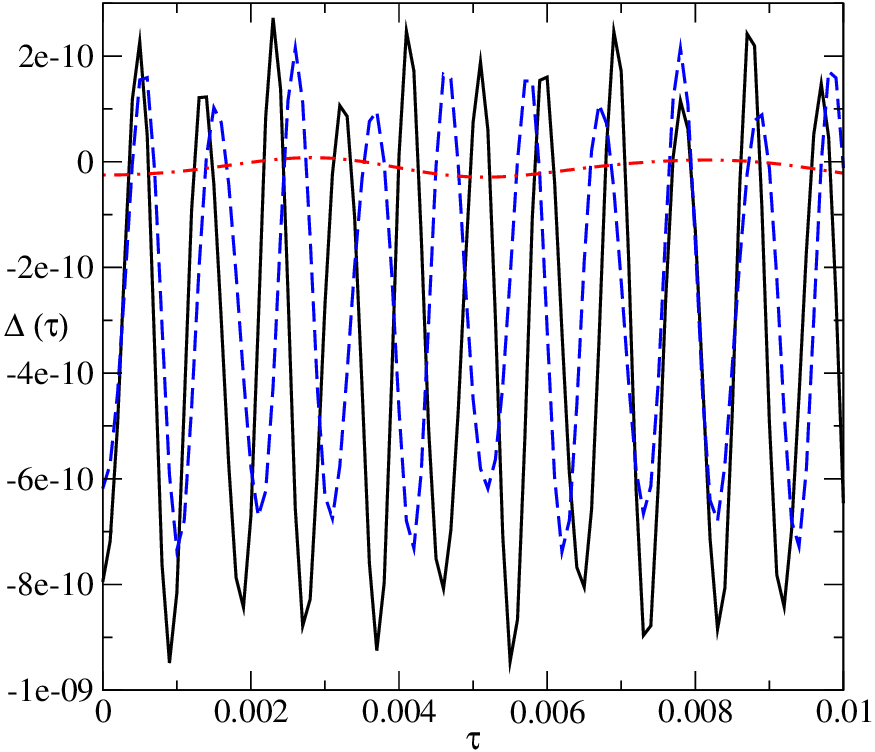}
	\vspace*{8pt}
	\caption{Normalized amplitude of the temperature oscillations of a gas in circular motion around electrically charged compact stars with mass $1.43 M_{\odot}$ (straight line), $1.765 M_{\odot}$ (dashed line) and $12.15 M_{\odot}$ (dot-dashed line). The proper time $\tau$ is expressed in seconds $(s)$.} 
	\label{fig01}
	
\end{figure}

\begin{figure}
	\includegraphics[width=6cm]{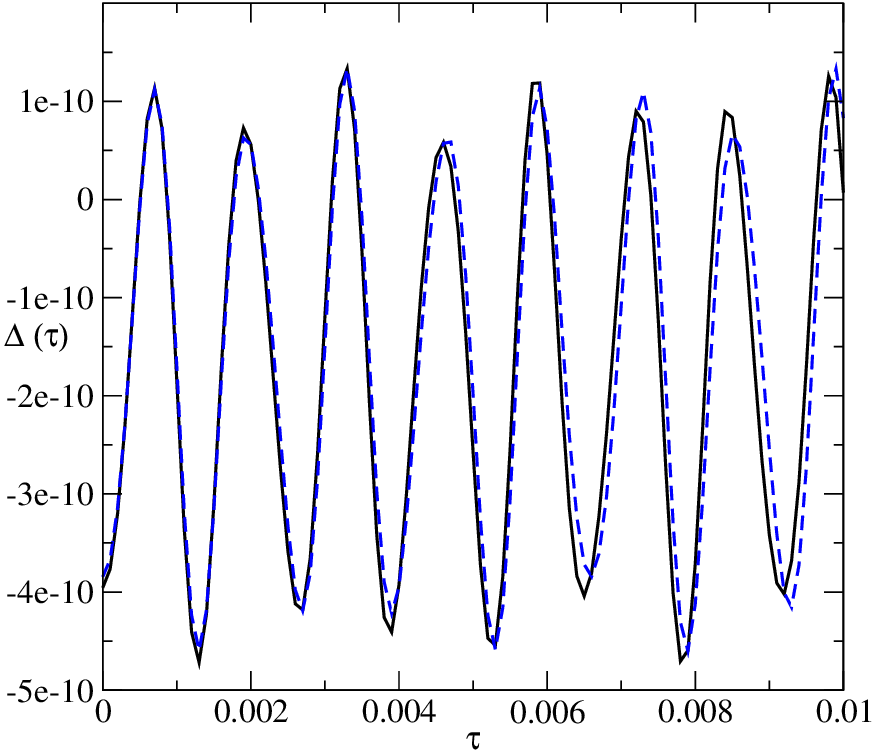}
	\vspace*{8pt}
	\caption{Normalized amplitude of the temperature oscillations of a gas in circular motion around electrically charged quark stars with mass $2.02 M_{\odot}$ (straight line) and $2.25 M_{\odot}$ (dashed line). The proper time $\tau$ is expressed in seconds $(s)$.}
	\label{fig02}       
\end{figure}

Using the values provided by the tables \ref{tab01} and \ref{tab02}, graphs relating the normalized amplitude of the temperature oscillations and proper time are plotted and shown in the figures \ref{fig01} and \ref{fig02}.
Comparing the plots shown in these figures  we can notice that the increase on the mass appears to affect the amplitude of temperature oscillations more than the increase on electric charge. It occurs due to the factor $M'/4r^3$ in the expression (\ref{eq167}). The radius is a multiple of the Reissner-Nordstr\"om radius, so it is also a function of the mass and the charge according with the positive root of the expression (\ref{eq45}). In our analysis we consider always $M > Q$ to get real roots, so the factor $M'/4r^3$ is $\mathcal{O}(M'^{-2})$. This is the reason why the increase of the mass causes a decrease in the amplitude of the temperature oscillations. However, if we keep the mass constant and increase the charge,  the amplitude tends to decrease because of the minus signal of the charge term in the root of expression (\ref{eq45}).
We also can notice that the curves shown in figure \ref{fig02} are much closer than the others in figure \ref{fig01}. This occurs because the increasing in charge and mass for the quark stars listed in table \ref{tab02} are small compared to the other star models. We can see that in the figure \ref{fig01}, the curves for $1.43 M_{\odot}$ and $1.765 M_{\odot}$ are closer from each other than the curve for $12.15 M_{\odot}$, because the increasing in mass and charge for the later is much larger than for the former ones.

\subsection{Charged black holes}
\label{sec03b}

In 2011 Bin-Nun \cite{binnun11} presented a model for describing electrically charged massive objects. This model considers the charge term in the Reissner-Nordstr\"om metric in terms of a proportion of the mass ($Q'=Q^2 = 4p_qM^2$, with $p_q$ as a free parameter).
According to this model, the quantity $Q'=Q^2$ of the Reissner-Nordstr\"om metric can be interpreted as a free parameter rather than a physical quantity such as electric charge. This interpretation considers the charge term as a result of a tidal gravitational effect \cite{dadhich00}, known as TRN ({\it Tidal Reissner-Nordstr\"om}). According to the TRN, the parameter $p_q$ can assume negative values.

A modification to this model was made in 2014 by Zakharov \cite{zakharov14}, expressing the free parameter $p_q$ as $p_q=Q^2/M^2$ without the $1/4 $ factor. This will be the expression adopted in the present analysis. Zakharov also presents some mass values for two black holes: one at the galactic center with mass $M_{BH} = (4,3 \pm 0,4) \times 10^6 M_{\odot}$ and another located in the elliptic galaxy $M87$ with mass $M_{M87} = (6,2 \pm 0,4) \times 10^9 M_{\odot}$. These values were obtained by observational measurements \cite{ghez08,gillessen09}.

The charge parameter $p_q$ assumes the values $0$ or $1$, where $p_q=0$ represents the absence of electric charge (Schwarzschild) and $p_q=1$ represents the case where $Q^2=M^2$, a situation known as ERN ({\it Extremal Reissner-Nordstr\"om}). The negative value of the charge parameter $p_q$ represents the TRN configuration.

Based on the article by Zakharov \cite{zakharov14} the tables \ref{tab03} and \ref{tab04} were elaborated, containing the mass, the different values of the charge parameter $p_q$, the orbit radius $r=5R_{RN}$ and the ratio between frequencies $\Omega/\omega$ for the two black holes considered in the article.

\begin{table}
	\caption{Charged black hole at the galactic center.}
	{\begin{tabular}{cccccc} 
			$M(M_{\odot})$ & $r = 5 R_{RN}(km)$ & $p_q $ & $\omega(mHz)$ & $\Omega(mHz)$& $\Omega/\omega$\\
			$4.3 \times 10^6$&$1.18 \times 10^8$&$-6.4$&$0.471$&$0.764$&$1.622$\\
			
			$4.3 \times 10^6$&$6.34 \times 10^7$&$0$&$1.131$&$1.789$&$1.581$\\
			
			$4.3 \times 10^6$&$3.17 \times 10^7$&$1$&$2.443$&$5.462$&$2.236$\\ 
		\end{tabular} \label{tab03}}
\end{table}

\begin{table}
	\caption{Charged black hole at M87.}
	{\begin{tabular}{cccccc} 
			$M(M_{\odot})$ & $r = 5 R_{RN}(km)$ & $p_q $ & $\omega(\mu Hz)$ & $\Omega(\mu Hz)$& $\Omega/\omega$\\
			$6.2 \times 10^9$&$1.70 \times 10^{11}$&$-6.4$&$0.326$&$0.529$&$1.622$\\
			
			$6.2 \times 10^9$&$9.14 \times 10^{10}$&$0$&$0.784$&$1.240$&$1.581$\\
			
			$6.2 \times 10^9$&$4.57 \times 10^{10}$&$1$&$1.694$&$3.788$&$2.236$\\ 
		\end{tabular} \label{tab04}}
\end{table}

\begin{figure}
	\includegraphics[width=6cm]{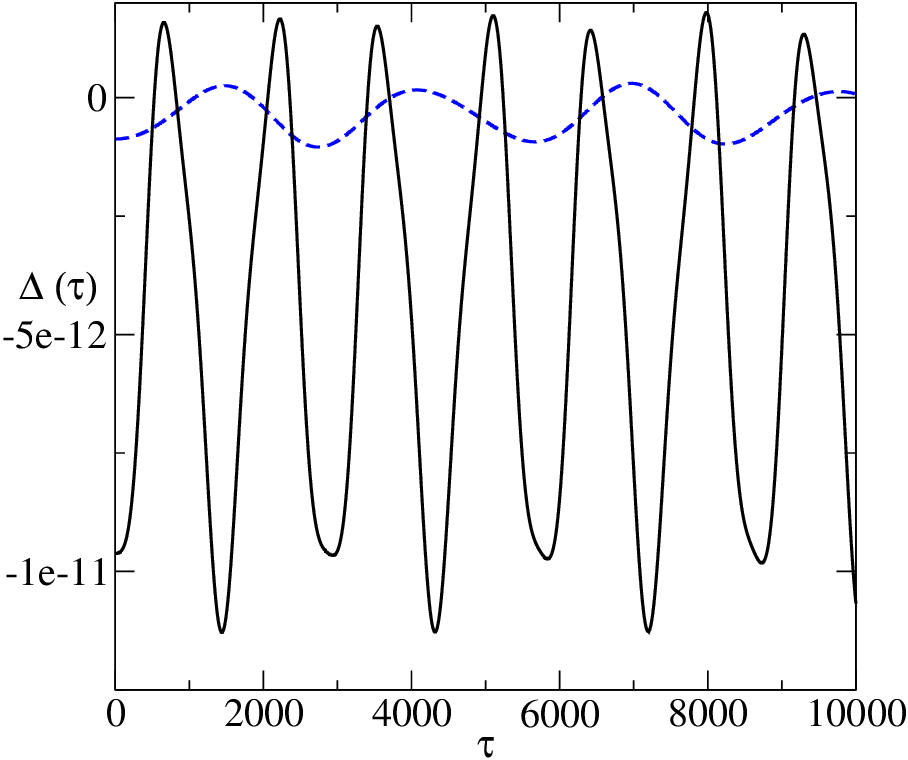}
	\vspace*{8pt}
	\caption{Normalized amplitude of the temperature oscillations of a gas in circular motion around a electrically charged black hole at the galactic center with charge parameter $p_q=0$ (dashed line) and $p_q=1$ (straight line). The amplitude $\Delta(\tau)$ is multiplied by a $10^{10}$ scale factor. The proper time $\tau$ is expressed in seconds $(s)$.}
	\label{fig03}       \end{figure}

\begin{figure}
	\includegraphics[width=6cm]{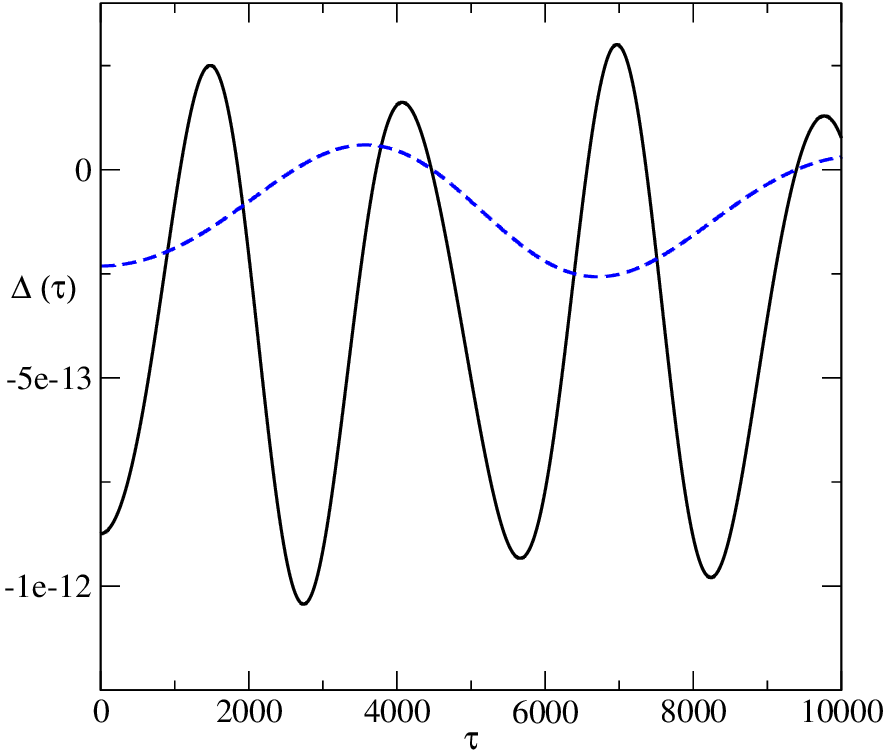}
	\vspace*{8pt}
	\caption{Normalized amplitude of the temperature oscillations of a gas in circular motion around a electrically charged black hole at the galactic center with charge parameter $p_q=0$ (straight line) and $p_q=-6.4$ (dashed line). The amplitude $\Delta(\tau)$ is multiplied by a $10^{10}$ scale factor. The proper time $\tau$ is expressed in seconds $(s)$.}
	\label{fig03b}      
\end{figure}

Using the values provided by tables \ref{tab03} and \ref{tab04}, graphs similar to the one shown in the figure \ref{fig03} have been plotted. The plot in \ref{fig03} relates the amplitude of temperature oscillations and the proper time for the situations where $p_q=0$ (Schwarzschild case) and $p_q=1$ (ERN case). The plot in \ref{fig03b}  relates the amplitude of temperature oscillations and the proper time for the cases where $p_q=0$ (Schwarzschild case) and $p_q=-6.4$ (TRN case). Comparing the plots shown in figures \ref{fig03} and \ref{fig03b} we can notice that the curve with charge parameter $p_q=-6.4$ has the smaller values for the amplitude of the oscillations. It occurs because of the factor $M'/4r^3$ in the expression (\ref{eq167}), where the radius is a multiple of the Reissner-Nordstr\"om radius. The negative value of the parameter $p_q$ causes the square root in expression (\ref{eq45}) always to have real roots, even if $M < Q$. So the factor $M'/4r^3$ is $\mathcal{O}(M'^{-2})$, with the charge contributing to increase the mass and subsequently decreasing the amplitude of the temperature oscillations.
We can also infer that the time scale in \ref{fig03} and \ref{fig03b} is much larger than the time scale for the other previous models. It is related with the values of the masses of the black holes, which are  much larger than the other compact objects. 

By analyzing the tables \ref{tab03} and \ref{tab04} we can realize that the values of the ratio $\Omega/\omega$ for the two black holes are the same for all values of $p_q$, even for different masses. This suggests that for this model, the ratio $\Omega/\omega$ does not depend on the mass, but only on the parameter $p_q$. In fact, we will show that for a model that considers the electric charge proportional to mass by a factor $p_q$, the ratio $\Omega/\omega$ depends only on $p_q$. To proof this assertion, let us build the ratio $\Omega/\omega$ from (\ref{s2}) and (\ref{s4})
by considering $Q^2 = p_q M'^2$ as in Bin-Nun's model, namely
\begin{equation}
\frac{\Omega}{\omega} = \sqrt{\frac{r^3 - r^2 p_q M'}{r^3-6M'r^2+9 p_q M'^2 r-4 p_q^2M'^3}}.
\label{eq173}
\end{equation}
Now by taking the orbit radius as a multiple of the Reissner-Nordstr\"om radius, we have:
\begin{gather}
r = N\left[M'(1+\sqrt{1-p_q})\right] = NM'h,
\label{eq174}
\end{gather}
where $N$ is a real positive number and $h = 1+\sqrt{1-p_q}$. Replacing this in (\ref{eq173}) we obtain:
\begin{gather}
\frac{\Omega}{\omega} = \sqrt{\frac{N^3 h^3 - N^2 h^2 p_q}{N^3 h^3-6 N^2 h^2 +9N p_q h-4 p_q^2}}.
\label{eq175}
\end{gather}
Therefore we conclude that $\Omega/\omega$ is only function of $p_q$.

\subsection{Charged white dwarfs}
\label{sec03c}

This model presented in 2014 by Liu et al \cite{liu14} suggests the possibility that electrically charged white dwarf stars are responsible for the formation of supernovae. These stars would have masses above the Chandrasekhar limit of $1,4 M_{\odot}$, so they would be unstable white dwarfs that would continue to collapse. In this model the polytropic equation of state  was considered for exact solutions, and a general equation of state (based on the equation of state of the free electron) was considered for numerical solutions.
For the present analysis the electric charge was considered with a constant value, with variable values of the mass. In this way we can see the role of the mass in the temperature oscillations.

Based on the values provided by the article of Liu et al \cite{liu14}, the table \ref{tab05} was elaborated, containing, in addition to mass and electric charge, the value of the orbit radius $r=5R_{RN}$ and the ratio between frequencies $\Omega/\omega$ of the white dwarf stars analyzed in the article.

\begin{table}
	\caption{Charged white dwarfs.}
	{\begin{tabular}{cccccc} 
			$M(M_{\odot})$ & $r = 5 R_{RN}(km)$ & $q(\times 10^{20} C)$ & $\omega(kHz)$ & $\Omega(kHz)$& $\Omega/\omega$\\
			$1.4325$&$13.622$&$2.35$&$5.899$&$10.559$&$1.789$\\
			
			$1.5168$&$15.967$&$2.35$&$4.955$&$8.460$&$1.707$\\
			
			$1.7970$&$21.816$&$2.35$&$3.502$&$5.719$&$1.633$\\
			
			$1.954$&$24.678$&$2.35$&$3.060$&$4.953$&$1.618$\\
			
			$2.4419$&$32.906$&$2.35$&$2.248$&$3.596$&$1.599$\\ 		\end{tabular} \label{tab05}}
\end{table}

\begin{figure}
	\includegraphics[width=6cm]{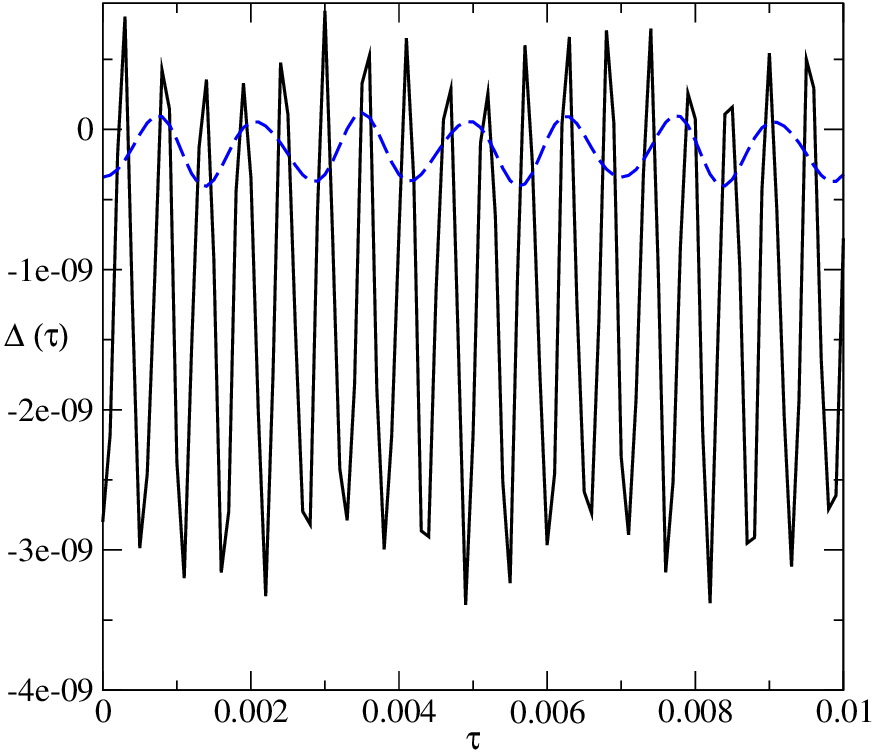}
	\vspace*{8pt}
	\caption{Normalized amplitude of the temperature oscillations of a gas in circular motion around a electrically charged white dwarf with mass $\approx 1.43 M_{\odot}$ (straight line) and $\approx 2.44 M_{\odot}$ (dashed line). The proper time $\tau$ is expressed in seconds $(s)$.}
	\label{fig04}       % Give a unique label
\end{figure}

Using the values provided by the table \ref{tab05} graphs similar to the one shown in the figure \ref{fig04} have been plotted. The plot in \ref{fig04} relates the amplitude of temperature oscillations and proper time for the mass values of $\approx 1,43 M_{\odot}$ and $\approx 2,44 M_{\odot}$.

\section{Summary and Conclusions}
\label{sec04}

In this work the normalized amplitudes of temperature oscillations of a gas in a circular geodesic motion in Reissner-Nordstr\"om space-time was determined by following the same  methodology of ref. \cite{zimdahl15} where  the Schwarzschild metric was used. The proper frequencies derived from the equations of the geodesic deviation in Reissner-Nordstr\"om metric were the same found by using the equations of motion. The expressions obtained for the temperature profile calculated from Tolman's law and the behavior of the oscillations were also consistent with the results of ref.\cite{zimdahl15}. 

After these calculations the obtained expression were used to calculate the amplitudes of temperature oscillations for some theoretical models for charged compact objects: compact stars, quark stars, black holes and white dwarfs. These oscillations presented frequencies of the same range $(kHz)$ as the frequencies found in QPO phenomena (with exception of the supermassive black holes, but the frequencies scale with the inverse of the mass and for this reason are in the $\mu Hz$ range \cite{kluzniak04}), so we calculate the ratio between the frequencies and compare with the $3/2$ ratio found in the literature about QPOs. Similar results were found, as for example the equilibrium tori model \cite{kovar11}. This model uses the equations of fluid mechanics to calculate the frequencies for some metrics \cite{abramowicz06}, including Reissner-Nordstr\"om, and the ratio $3/2$ was also obtained \cite{bursa04}. 

For the compact and quark star models, it has been found that the $\Omega/\omega$ ratio is closer to $3/2$ for the configurations with the lowest values of charge and mass. It was also shown that small variations in mass can compensate for large variations in electric charge, as observed in the quark star model.

The model for black holes presented mass and charge values many orders of magnitude higher than the other compact objects analyzed, consequently the amplitude values were smaller by many orders of magnitude. To allow comparison with the results obtained from the other models, it was necessary to multiply the amplitudes by a scale factor. Although any value of charge and mass could be chosen for our analysis, it was preferred to use values from a model found in the literature. For this model, where the electric charge was taken proportional to the mass, it was possible to show that the ratio $\Omega/\omega$ does not depend on mass but only on the proportionality factor between charge and mass. We can also observe that the ratio $\Omega/\omega$ increases with the electric charge. For the situation with charge parameter $p_q = 1$, this value is much larger than $3/2$, but this is an extreme limiting case.   

The white dwarf model allowed to analyze the variation of mass with constant electric charge. Comparing the analysis of this model with that of the black holes, it was possible to verify that the role of the charge term is the opposite of the mass term, that is, while the increase of mass produces a reduction in the frequencies, amplitude and in the ratio between frequencies, the increase of the electric charge produces an inverse effect. This behavior reflects the fact that the mass term and the electric charge term have opposite signs in the expression of the Reissner-Nordstr\"om metric. In this way, the main objective of the work was reached, which was the determination of the role of the charge term in the behavior of temperature oscillations of a gas in geodesic motion in the presence of a Reissner-Nordstr\"om metric.

Although the QPOs and the problem treated in the present work are not the same problem, they have some features in common. The study of a gas in circular geodesic motion around a charged compact object based in the Lagrangian of a free particle motion reveals the existence of two proper frequencies: $\omega$ associated with the radial frequency $\omega_r$ and $\Omega$ associated with the orbital frequency $\omega_\phi$. The existence of these two frequencies and the relativistic strong field regime suggests that these two problems can share some common behaviors. For these reasons we also calculate the ratio between the frequencies $\Omega / \omega$ in our analysis. As said in the introdution, this is an idealized toy model with several limitations but some results are compatible with the values found for QPOs.

\section*{Acknowledgments}

L. C. M. has been supported by CAPES (Coordena\c c\~ao de Aperfei\c coamento de Pessoal de N\'ivel Superior), Brazil and G. M. K by  CNPq (Conselho Nacional de Desenvolvimento Cient\'ifico e Tecnol\'ogico), Brazil. The authors thank Professor Winfried Zimdahl for useful discussions.

%\begin{thebibliography}{000} %for 3 digits
%\begin{thebibliography}{00}  %for 2 digits


\begin{thebibliography}{0}    %for 1 digit

\bibitem{wald} R. M. Wald, {\it General Relativity}, (The University of Chicago Press, Chicago, 1984).

\bibitem{misner} C. Misner, K. Thorne, and J. Wheeler, {\it Gravitation}, (W.H. Freedman and
Company, San Francisco, 1973).

\bibitem{chandrasekhar} S. Chandrasekhar, {\it The Mathematical Theory of Black Holes}, (Oxford University
Press, Oxford, 2009).

\bibitem{ray03} S. Ray et al., {\it Phys. Rev. D}, {\bf 68} (2003) 84004.

\bibitem{binnun11} A. Bin-Nun, {\it Class. Quantum Grav.}, {\bf 28} (2011) 114003.

\bibitem{zakharov14} A. Zakharov, {\it Phys. Rev. D}, {\bf 90} (2014) 62007.

\bibitem{liu14} H. Liu et al., {\it Phys. Rev. D}, {\bf 89} (2014) 104043.

\bibitem{zimdahl15} W. Zimdahl and G. M. Kremer, {\it Phys. Rev. D}, {\bf 91} (2015) 24003.

\bibitem{negreiros09} R. Negreiros et al., {\it Physical Review D},
{\bf 80} (2009) 83006.

\bibitem{cercignani} C. Cercignani and G. M. Kremer, {\it The Relativistic Boltzmann Equation: Theory and Applications}, (Birkh\"auser Verlag, Berlin, 2002).

\bibitem{tolman30} R. Tolman and P. Ehrenfest, {\it Phys. Rev.}, {\bf 36} (1930) 1791.

\bibitem{kovar11} J. Kov\'a\v{r} et al., {\it Phys. Rev. D}, {\bf 84} (2011) 84002.

\bibitem{klein} O. Klein, {\it Rev. Mod. Phys.}, {\bf 21} (1949) 531.

\bibitem{hees17} A. Hees et al., {\it Phys. Rev. Lett.}, {\bf 118} (2017) 211101.

\bibitem{broderick13} A. Broderick et al., {\it Astrophys. J.}, vol. {\bf 784} (2014) 7.

\bibitem{zhang15} F. Zhang, Y. Lu and Q. Yu, {\it Astrophys. J.}, {\bf 809} (2015) 27.

\bibitem{parker82} L. Parker and L. Pimentel, {\it Phys. Rev. D}, {\bf 25} (1982) 3180.

\bibitem{shirokov73} M. Shirokov, {\it Gen. Relativ. Gravit.}, {\bf 4} (1973) 131.

\bibitem{zimdahl85} W. Zimdahl, {\it Exp. Tech. Phys.}, {\bf 33} (1985) 403.

\bibitem{manasse63} F. Manasse and C. Misner, {\it J. Math. Phys.}, {\bf 4} (1963) 735.

\bibitem{torok05} G. T\"or\"ok et al., {\it Astron. Astrophys.}, {\bf 436} (2005) 1.

\bibitem{germana13} C.German\'a, {\it Mon. Not. R. Astron. Soc.}, {\bf 430} (2013) L1.

\bibitem{kluzniak01} W. Klu\'zniak and M. Abramowicz, {\it Astron. Astrophys.}, {\bf 374} (2001) L19.

\bibitem{kluzniak04} M. Abramowicz et al., {\it Astrophys. J.}, {\bf 609} (2004) L63.

\bibitem{dinverno} R. D'Inverno, {\it Introducing Einstein's Relativity}, (Oxford University Press, Oxford, 1998).

\bibitem{collas07} P. Collas and D. Klein, {\it Gen. Rel. Grav.}, {\bf 39} (2007) 737.

\bibitem{dadhich00} N. Dadhich et al., {\it Phys. Lett. B}, {\bf 487} (2000) 1.

\bibitem{ghez08} A. Ghez et al., {\it Astrophys. J.}, {\bf 689} (2008) 1044.

\bibitem{gillessen09} S. Gillessen et al., {\it Astrophys. J.}, {\bf 707} (2009) L114.



\bibitem{abramowicz06} M. Abramowicz et al., {\it Class. Quantum Grav.}, {\bf 23} (2006) 1689.

\bibitem{bursa04} M. Bursa et al., {\it Astrophys. J.}, {\bf 617} (2004) L45.


\end{thebibliography}
\end{document}